\documentclass[conference]{IEEEtran}
\usepackage{etoolbox}
\makeatletter
\patchcmd{\@makecaption}
{\\}
{.\ }
{}
{}
\makeatother

\usepackage{booktabs}
\usepackage{tikz}
\usepackage{url}
\usepackage{pdfpages}
\usepackage{afterpage}
\usepackage{color, colortbl}
\usepackage{graphicx}
\usepackage{amsmath}
\usepackage{import}
\usepackage{pifont}
\usepackage{multirow}
\usepackage{subfigure}
\usepackage{footmisc}
\usepackage{algorithm}
\usepackage{lipsum}

\usepackage[
separate-uncertainty = true,
multi-part-units = repeat
]{siunitx}
\usepackage{algpseudocode}

\usepackage{array}
\usepackage{url}
\newcommand\Mark[1]{\textsuperscript#1}

\def\BState{\State\hskip-\ALG@thistlm}

\definecolor{Gray}{gray}{0.9}
\definecolor{LightCyan}{rgb}{0.88,1,1}
\pdfoutput=1
\graphicspath {{figures/}}

\usepackage[flushleft]{threeparttable}

\begin{document}

\title{A DAG Model of Synchronous Stochastic Gradient Descent in Distributed Deep Learning}

%\author{\Mark{\dag}Shaohuai Shi, \Mark{\dag}Qiang Wang, \Mark{\dag}Xiaowen Chu, \Mark{\ddag}Bo Li}
%\affiliation{%
%	\depthof{\Mark{\small\dag}Department of Computer Science, Hong Kong Baptist University} \\
%	\institution{\Mark{\small\ddag}Department of Computer Science and Engineering, The Hong Kong University of Science and Technology}
%	%  \city{Hong Kong}
%	%  \state{China}
%	\city{\Mark{\dag}\{csshshi,qiangwang,chxw\}@comp.hkbu.edu.hk, \Mark{\ddag}bli@cse.ust.hk}
%}

\author{\IEEEauthorblockN{\Mark{\dag}Shaohuai Shi, \Mark{\dag}Qiang Wang, \Mark{\dag}Xiaowen Chu, \Mark{\small\ddag}Bo Li}
	\IEEEauthorblockA{\Mark{\dag}Department of Computer Science, Hong Kong Baptist University
		\\\Mark{\small\ddag}Department of Computer Science and Engineering, The Hong Kong University of Science and Technology
		\\\Mark{\dag}\{csshshi, qiangwang, chxw\}@comp.hkbu.edu.hk,\Mark{\ddag}bli@cse.ust.hk}
}
%\author{\IEEEauthorblockN{Bo Li}
%	\IEEEauthorblockA{The Hong Kong University of Science and Technology
%		\\bli@cse.ust.hk}
%}

\maketitle
%\author{Shaohuai Shi}
%\affiliation{%
%	\depthof{Department of Computer Science, }
%	\institution{Hong Kong Baptist University}
%	%  \city{Hong Kong}
%	%  \state{China}
%}
%\email{csshshi@comp.hkbu.edu.hk}
%\author{Qiang Wang}
%\affiliation{%
%	\depthof{Department of Computer Science, }
%	\institution{Hong Kong Baptist University}
%	%  \city{Hong Kong}
%	%  \state{China}
%}
%\email{qiangwang@comp.hkbu.edu.hk}
%\author{Xiaowen Chu}
%\affiliation{%
%	\depthof{Department of Computer Science, }
%  \institution{Hong Kong Baptist University}
%%  \city{Hong Kong}
%%  \state{China}
%}
%\email{chxw@comp.hkbu.edu.hk}
%\author{Bo Li}
%\affiliation{%
%	\depthof{Department of Computer Science and Engineering, }
%	\institution{The Hong Kong University of Science and Technology}
%	%  \city{Hong Kong}
%	%  \state{China}
%}
%\email{bli@cse.ust.hk}

\begin{abstract}
With huge amounts of training data, deep learning has made great breakthroughs in many artificial intelligence (AI) applications. However, such large-scale data sets present computational challenges, requiring training to be distributed on a cluster equipped with accelerators like GPUs. With the fast increase of GPU computing power, the data communications among GPUs have become a potential bottleneck on the overall training performance. In this paper, we first propose a general directed acyclic graph (DAG) model to describe the distributed synchronous stochastic gradient descent (S-SGD) algorithm, which has been widely used in distributed deep learning frameworks. To understand the practical impact of data communications on training performance, we conduct extensive empirical studies on four state-of-the-art distributed deep learning frameworks (i.e., Caffe-MPI, CNTK, MXNet and TensorFlow) over multi-GPU and multi-node environments with different data communication techniques, including PCIe, NVLink, 10GbE, and InfiniBand. Through both analytical and experimental studies, we identify the potential bottlenecks and overheads that could be further optimized. At last, we make the data set of our experimental traces publicly available, which could be used to support simulation-based studies.
\end{abstract}


\IEEEpeerreviewmaketitle

\section{Introduction}
Recently, deep learning (DL) techniques have achieved great success in many AI applications \cite{lecun2015deep} such as image classification, speech recognition and generation, and natural language processing. Deep model training aims to learn the set of model parameters by minimizing a loss function iteratively \cite{bottou2010large}\cite{wang2017stochastic}. With increases in the training data size and the complexity of deep models, it becomes necessary to scale out the time-consuming training jobs to GPU clusters through either model parallelization \cite{lee2014model} or data parallelization \cite{zinkevich2010parallelized}\cite{you2017scaling}\cite{goyal2017accurate}\cite{raina2009large}\cite{chen2016revisiting}. This pushes the technology giants to deploy their cloud-based AI services with highly scalable deep learning tools. For example, Amazon adopts MXNet \cite{chen2015mxnet} as the main DL framework for cloud service AWS; Google develops TensorFlow \cite{abadi2015tensorflow} for Google Cloud; and Microsoft develops Microsoft Cognitive Toolkit (CNTK) \cite{seide2016cntk} for Microsoft Azure.

When training a deep model on distributed systems, the computing tasks are typically carried out by a set of worker GPUs in many iterations. Each iteration includes two major steps: feed-forward and back propagation. Within each iteration, the GPUs need to exchange huge amount of information about the model parameters or gradients. Most of the existing work focuses on the optimization of the computing primitives in deep learning, such as dense and sparse matrix multiplication, convolution, FFT, etc. With the advance of GPU computing power and newly designed parallel algorithms (e.g., cuDNN \cite{chetlur2014cudnn}), the computing time of feed-forward and back propagation has been significantly reduced, making the data communication overhead a potential system bottleneck. Some existing empirical studies have shown that the current deep learning frameworks do not scale very well on GPU clusters due to the overhead of data communications \cite{bahrampour2015comparative}\cite{shi2016benchmarking}\cite{shams2017evaluation}\cite{kim2017performance}, but there lacks an in-depth study of the impact of data communications on the overall training performance.

In this paper, we try to investigate the impact of data communications on distributed training with the synchronous stochastic gradient descent (S-SGD) algorithm, which has been widely used by mainstream deep learning frameworks. We first propose a general directed acyclic graph (DAG) model to describe the S-SGD algorithm. Different from the traditional DAG model whose nodes represent computing tasks, our DAG model includes two types of nodes: computing and communication nodes. The edges are used to describe the precedence constraint between two tasks. We use the DAG model to explore the possible strategies of reducing the training time, and compare four distributed DL frameworks (i.e., Caffe-MPI\cite{caffempi}, CNTK, MXNet and TensorFlow). We then conduct empirical studies on these DL framework, aiming at understanding how different data communication techniques (e.g., PCIe, NVLink, 10GbE and InfiniBand) and optimization strategies affect the training performance. Our major findings are summarized as follows:
\begin{enumerate}
    \item The different implementations of the S-SGD algorithm on all studied frameworks can be well described by our general DAG model. The speedup depends on three major factors: I/O performance, computing performance, and communication performance.
    \item Through the DAG model, we show two optimization opportunities: overlapping I/O with computing, and overlapping gradient aggregation with computing. In S-SGD with multiple GPUs, CNTK does not hide the overhead of gradient communication, while Caffe-MPI, MXNet and TensorFlow parallelize the gradient aggregation with the gradient computation. By hiding the overhead of gradient communication, the scaling performance could be improved.
	\item All the frameworks scale not very well on the most advanced GPUs (Nvidia Tesla V100) when Tensor Cores are utilized. The current implementations of inter-node gradient communication via 100Gbps InfiniBand are still not good enough to match the computing power of V100.
    \item On certain deep neural networks, we observe very low utilization of network bandwidth on InfiniBand due to the layer-wise pattern of data communications during back propagation.
\end{enumerate}

The rest of the paper is organized as follows. Section \ref{backgroundandrelatedwork} presents some related work. Section \ref{preliminaries} introduces the SGD and S-SGD algorithms. We propose our general DAG model and discuss different performance optimization strategies in Section \ref{modeling}. Our experiments and analysis are presented in Section \ref{methods}, followed by the introduction to the published layer-wise trace data set in Section \ref{dataset}. We conclude the paper in Section \ref{conclusionandfuturework}.

\section{Related Work} \label{backgroundandrelatedwork}
Data parallelism synchronous SGD (S-SGD) is widely used in the training of deep neural networks to scale to multiple GPUs or machines without affecting the convergence performance \cite{das2016distributed}\cite{chen2016revisiting}\cite{wang2017stochastic}\cite{you2017scaling}\cite{goyal2017accurate}\cite{you2017100}, and hence becomes the embedded component in mainstream DL frameworks such as Caffe-MPI, CNTK, MXNet and TensorFlow. However, because of the different design philosophy of software by vendors, these frameworks implement S-SGD differently so that the scaling performance varies. Bahrampour et al. \cite{bahrampour2015comparative} and Shi et al. \cite{shi2016benchmarking} have evaluated the performance of some DL frameworks on a GPU. In the distributed environment, Shams et al. \cite{shams2017evaluation} have studied the performance of Nvidia's NVLink and Intel\rq s Knights Landing on different CPU and GPU technologies. However, some other popular DL frameworks (e.g., Caffe-MPI, CNTK and MXNet) are not evaluated in \cite{shams2017evaluation}. In addition, there lacks an in-depth analysis of the scalability performance of S-SGD algorithm in distributed clusters. Shi et al., \cite{shi2017performance} evaluate the same distributed frameworks with high-speed networks, but they do not provide a comparison between slow- and high- speed connections. The main differences of this work compared to the work in \cite{shi2017performance} are that we build a directed acyclic graph to generalize the performance of S-SGD and compare the impact of the high- and slow- speed networks on training DNNs. 

S-SGD requires the set of computing units (e.g., GPUs) to exchange data iteratively, which can be implemented by either parameter server (PS) based methods \cite{li2014scaling}\cite{cui2016geeps} or decentralized methods. In PS-based methods, there is one or more PSes that store the global model. The PS aggregates parameters at each iteration, updates the model, and then pushes the updated model to each computing unit. Performance models have been built by S. Zou et al. \cite{zou2017distributed} to generalize the performance of the PS-based methods, which provides guidelines for better system scalability.

Decentralized methods implement the gradients aggregation by using the reduction tree (RT) or ring based all-reduce \cite{turchenko2010improvement}\cite{awan2016efficient}\cite{awan2017s}. The gradients are exchanged via MPI-like collectives (e.g., all-reduce). Very recently, some new collective communications libraries like Gloo\footnote{https://github.com/facebookincubator/gloo} and NCCL2\footnote{https://developer.nvidia.com/nccl} have been developed to support efficient communications among a set of GPUs. A. Awan et al. \cite{awan2017s}\cite{awan2017optimized} propose a high performance CUDA-Aware MPI to reduce the overhead of data communications across a GPU cluster. He et al. have shown that the optimized all-reduce implementation and the pipeline of all-reduce operations with gradient computation can lead to very good scalability \cite{he2015deep}.

Different from the above studies, in this paper we first propose a DAG model to generalize the workflow of distributed training with S-SGD, and then study several state-of-the-art deep learning frameworks under multi-GPU and multi-node environments through theoretical analysis and real-world experiments.

\section{SGD and S-SGD} \label{preliminaries}
In this section, the algorithms of SGD and S-SGD are introduced. For easy reference, some mathematical notations are summarized in Table \ref{table:notation}.

\begin{table}[!ht]
	\centering
	\caption{Summary of notations}
	\label{table:notation}
	\begin{tabular}{|l|l|}
		\hline
		Name &  Description \\\cline{1-2}
		\hline
		\hline
		$N$ & \# of machines in the cluster \\\cline{1-2}
		$n_g$ & \# of GPUs on each node \\\cline{1-2}
		$N_g$ & \# of total GPUs, $N_g=N\times n_g$ \\\cline{1-2}
		$M$ & \# of training samples per GPU in a mini-batch \\\cline{1-2}
		$L$ & The number of layers in a deep neural network\\\cline{1-2}
		$t_{iter}$ & Time of an iteration\\\cline{1-2}
		$t_{io}$ & Time of I/O in each iteration\\\cline{1-2}
		$t_{h2d}$ & Data transfer time from CPU memory to GPU memory in each iteration\\\cline{1-2}
		$t_{f}$ & Time of the forward phase in each iteration\\\cline{1-2}
		$t_{f}^{(l)}$ & Time of the forward phase of layer $l$ in each iteration\\\cline{1-2}
		$t_{b}$ & Time of the backward phase in each iteration\\\cline{1-2}
		$t_{b}^{(l)}$ & Time of the backward phase of layer $l$ in each iteration\\\cline{1-2}
		$t_{u}$ & Time of the model update in each iteration\\\cline{1-2}
		$t_{c}$ & Time of the gradients aggregation in each iteration\\\cline{1-2}
		$t_{c}^{(l)}$ & Gradients aggregation time of layer $l$ in each iteration\\\cline{1-2}
		$t_{c}^{no}$ & Time of non-overlapped gradient communications \\\cline{1-2}
%		$C$ & The assumption of the $C^{th}$ learnable layer, \\
%		&$t_{comm}^{(i)} \leq t_{b}^{(i-1)}$ for $i=2,3,..., C-1$, and\\
%		&$t_{comm}^{(i)} > t_{b}^{(i-1)}$ for $i=C,C+1,...,L$ \\\cline{1-2}	
	\end{tabular}
%	\vspace{-10pt}
\end{table}

\subsection{Mini-batch SGD} \label{sgd}
Assume that an $L$-layer model is trained with the mini-batch SGD on a GPU, the layer-wise parameters of the model are updated iteratively. Each iteration generally contains five steps: 1) Fetch data: Load a mini-batch of training data from the disk or the cache; 2) Data transfer through PCIe: Transfer the training data from CPU memory to GPU memory; 3) Feed-forward: Use GPU to perform feed-forward calculations from layer $1$ to layer $L$; 4) Back propagation: Use GPU to calculate gradients from layer $L$ back to layer $1$; 5) Update: The model are updated by the calculated gradients in the previous step. The time of one iteration can be represented by
\begin{equation}
t_{iter}=t_{io}+t_{h2d} + t_{f} + t_{b} + t_{u}=t_{io}+t_{h2d} + \sum_{l=1}^{L}t_{f}^{(l)}+\sum_{l=L}^{1}t_{b}^{(l)}+ t_{u}.
\end{equation}

\subsection{S-SGD on multiple GPUs} \label{ssgd}
The pseudo-code of S-SGD is shown in Algorithm \ref{algo:ssgd}. S-SGD makes each GPU perform feed-forward and backward propagation in parallel with different training data on the same model. Compared to SGD, S-SGD contains six steps, and the first four steps are the same with SGD (i.e., Fetch data, data transfer through PCIe, feed-forward and back propagation). The fifth step is an extra operation (i.e., gradient aggregation) before udpating the model in the sixth step. The iteration time $t_{iter}$ for the naive S-SGD implementation can be represented as:
\begin{equation}\label{seqiter}
t_{iter} = t_{io} + t_{h2d} + \sum_{l=1}^{L}t_{f}^{(l)}+\sum_{l=1}^{L}t_{b}^{(l)}+ \sum_{l=1}^{L}t_{c}^{(l)} + t_{u}.
\end{equation}
In the single-GPU environment, $t_{c}^{(l)}=0$. 

\begin{algorithm}
	\caption{S-SGD}\label{algo:ssgd}
	\begin{algorithmic}[1]
		\Procedure{S-SGD}{parameters, data, $N$}
		\For{$\textit{each worker i} \in \{1, 2, ..., N\}$}
		\State FeedForward(parameters, $\frac{data}{N}$)
		\State $\nabla{g}_i \gets \text{BackPropagation}()$
		\EndFor
		\State Synchronous()
		\State $\textit{Aggregate from all workers: } \nabla{g} \gets \frac{1}{N}\sum_{i=1}^{N}{\nabla{g}_i}$
		\For{$\textit{each worker i} \in \{1, 2, ..., N\}$}
			\State UpdateModel()
		\EndFor
		\EndProcedure
	\end{algorithmic}
\end{algorithm}

\section{A DAG Model of S-SGD} \label{modeling}
In this section, we first define two types of tasks in distributed training of deep neural networks, and then propose a general directed acyclic graph (DAG) model for S-SGD based training. After that, we apply the DAG model to discuss different optimization strategies in Caffe-MPI, CNTK, MXNet and TensorFlow.

\subsection{Definitions}
We define the following two types of tasks in a training job:

\begin{itemize}
	\item \textbf{Computing task}, whose resource requirement is mainly on the computational units (e.g., CPUs and GPUs).
	\item \textbf{Communication task}, whose resource requirement is the disk I/O or the interconnect (e.g., PCIe, NVLink, Ethernet, and InfiniBand) between computing units.
\end{itemize}

Considering S-SGD, in the first step, each GPU needs $M$ data samples separately, and the data should be read from the disk, so fetching $M$ samples can be regarded as a communication task. In the second step, the fetched data should be transferred to GPUs via PCIe, so we regard each transmission of $M$ samples between CPU memory and GPU memory as a communication task. In the third and fourth steps, the data is fed-forward on GPU layer by layer, and then back propagated layer by layer, both of which require GPU to carry out calculations. So each layer's feed-forward can be regarded as a computing task, and so does each layer's back propagation. In the fifth step, the gradients of each layer should be aggregated by all GPUs via intra-connect (e.g., PCIe and NVLink) and/or inter-connect (e.g., Ethernet and InfiniBand) communications, which is regarded as a communication task.

\subsection{A DAG model of S-SGD}
Directed acyclic graph (DAG) is a popular approach to modeling complex computing jobs, in which the nodes represent the computing tasks and the edges represent the precedence constraint between two tasks. In this paper, we focus on distributed training jobs that involve extensive data communications. To this end, we introduce communication nodes into the DAG to represent the communication tasks throughout the training job. To be more specific, a job $J$ is represented by a DAG $G = (V_c\bigcup V_n, E)$, where $V_c$, $V_n$, $E$ are the set of computing nodes (or tasks), communication nodes (or tasks), and directed edges, respectively. A directed edge $e_{x,y}$ from node $x$ to node $y$ represents the precedence constraint that task $y$ can only begin after task $x$ is finished.

\begin{figure*}[!ht]
	\centering
	\includegraphics[width=0.745\linewidth]{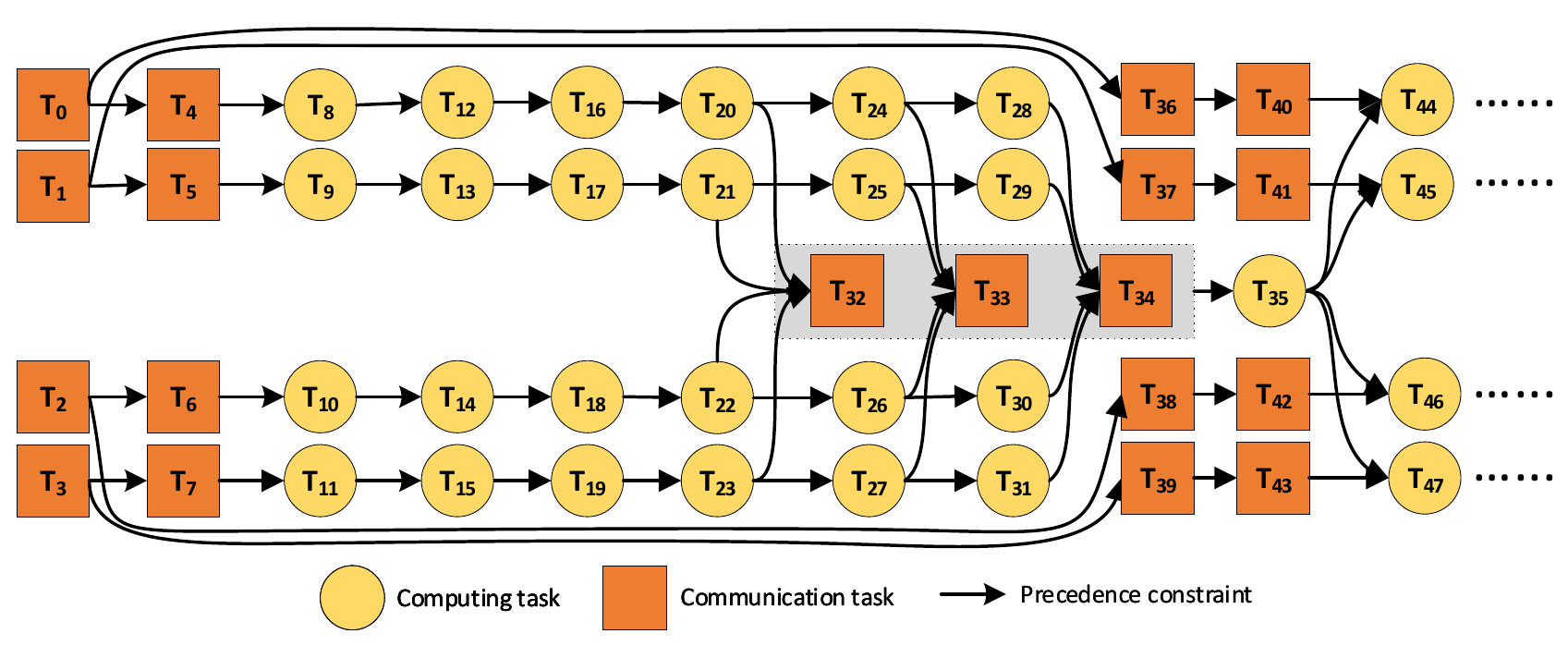}
	\caption{A DAG model of training a 3-layer neural network with 4 GPUs.}
	\label{fig:dag}
%		\vspace{-10pt}
\end{figure*}

A DAG example of a distributed training job using S-SGD is shown in Fig. \ref{fig:dag}. The training job is to train a 3-layer model using 4 GPUs. The yellow circle nodes represent the computing tasks; the orange square nodes represent the communication tasks; and the directed edges represent the precedence constraint between two tasks. Tasks $\mathsf{T}_0$-$\mathsf{T}_3$ read the training data from the disk or the network file system, which are classified as communication tasks. Tasks $\mathsf{T}_4$-$\mathsf{T}_7$ transfer data from CPU memory to GPU memory, which are also classified as communication tasks. Tasks $\mathsf{T}_8$-$\mathsf{T}_{11}$ represent the feed-forward computing tasks of layer 1, followed by Tasks $\mathsf{T}_{12}$-$\mathsf{T}_{15}$ for layer 2 and Tasks $\mathsf{T}_{16}$-$\mathsf{T}_{19}$ for layer 3. Tasks $\mathsf{T}_{20}$-$\mathsf{T}_{23}$ represent the back propagation computing tasks of layer 3, followed by Tasks $\mathsf{T}_{24}$-$\mathsf{T}_{27}$ for layer 2 and Tasks $\mathsf{T}_{28}$-$\mathsf{T}_{31}$ for layer 1. Tasks $\mathsf{T}_{32}$, $\mathsf{T}_{33}$ and $\mathsf{T}_{34}$ aggregate the gradients of layer 3, 2 and 1, respectively, which can be implemented by all-reduce communication tasks. Task $\mathsf{T}_{35}$ updates the model, which is a computing task that depends on $\mathsf{T}_{32}$, $\mathsf{T}_{33}$ and $\mathsf{T}_{34}$. Tasks $\mathsf{T}_{36}$-$\mathsf{T}_{39}$ represent the loading of another mini-batch of data for the next iteration.

\subsection{Optimization opportunities}
According to the proposed DAG model for S-SGD, the computing tasks of a new iteration cannot begin before the model update task of the previous iteration. We can observe two possible optimization opportunities with pipeline techniques. The first one is to parallelize the tasks of data reading (e.g., tasks $\mathsf{T}_{36}$-$\mathsf{T}_{39}$) with the computing tasks (e.g., tasks $\mathsf{T}_{8}$-$\mathsf{T}_{31}$), which could hide the time cost of disk I/O. The second one is to parallelize the gradient communication tasks (e.g., tasks $\mathsf{T}_{32}$-$\mathsf{T}_{34}$) with the back propagation computing tasks (e.g., tasks $\mathsf{T}_{24}$-$\mathsf{T}_{31}$).

\textbf{Overlapping I/O with computing}. The tasks of data fetching for an iteration has no edge connections with the computing tasks of its previous iteration, so one can parallelize these two types of tasks such that the I/O time can overlap with the computing tasks. Taking the example of Fig. \ref{fig:dag}, tasks $\mathsf{T}_{36}$-$\mathsf{T}_{39}$ can immediately begin after tasks $\mathsf{T}_{0}$-$\mathsf{T}_{3}$ have finished, and then be followed by the communication tasks $\mathsf{T}_{40}$-$\mathsf{T}_{43}$. In such a way, computing tasks $\mathsf{T}_{44}$-$\mathsf{T}_{47}$ in the next iteration can begin immediately after $\mathsf{T}_{35}$ is finished. The average iteration time when we overlap I/O with computing can be estimated by
\begin{equation}
\label{equ:pipeioiter}
\bar{t}_{iter} = \text{max}\{t_{io}+t_{h2d}, t_{f} + t_{b} + t_{c}\}.
\end{equation}
Notice that if the tasks of data transfer from CPU memory to GPU memory begin immediately after the data fetching is finished, the system requires extra GPU memory for holding new training data.
%\subsection{Communication via NCCL2}
%Gradients aggregation in S-SGD also easily becomes the bottleneck of performance due to the communication overhead via high-latency devices (i.e., PCIe and network interfaces). NCCL2 is a high-performance library to exchange data across multiple GPUs both in the single node and the distributed environment. Therefore, it is somehow effective to reduce the value of $t_{comm}$ by using NCCL2.

\textbf{Overlapping gradient communication with computing}. The gradient communication tasks can also be parallelized with the back propagation computing tasks. For instance, the communication task $\mathsf{T}_{32}$ can be parallelized with computing tasks of $\mathsf{T}_{24}$-$\mathsf{T}_{27}$, and the communication task $\mathsf{T}_{33}$ can be parallelized with computing tasks of $\mathsf{T}_{28}$-$\mathsf{T}_{31}$. This strategy is also known as the wait-free back-propagation (WFBP) algorithm \cite{zhang2017poseidon}\cite{awan2017s}. Let $\tau_{c}^{(l)}$ and $\mu_{c}^{(l)}$ denote the start and the end time of gradient communication of layer $l$ during one iteration, respectively. Then the iteration time is
\begin{equation}
\bar{t}_{iter} = \text{max}\{t_{io} + t_{h2d}, t_{f} + t_{b}^{(L)} + \tau_{c}^{(1)} - \tau_{c}^{(L)} + t_{c}^{(1)}\},
\end{equation}
where $t_{b}^{(L)}$ is the gradient computation time of the last layer. Since the gradient communication task of layer $l-1$ depends on the gradient computing task of layer $l-1$, we have $\tau_{c}^{(1)} \geq \tau_{b}^{(1)}+t_{b}^{(1)}$ and $\tau_{c}^{(L)}=\tau_{b}^{L}+t_{b}^{(L)}$, where $\tau_{b}^{(l)}$ is the start time of the gradient computing task of layer $l$. We use $t_{c}^{no}$ to denote the non-overlapped time cost of gradient communication tasks. Thus, we have
\begin{equation}
\bar{t}_{iter} = \text{max}\{t_{io} + t_{h2d}, t_{f} + t_{b} + t_{c}^{no}\}.
\end{equation}

Let $t_{iter\_x}$ and $t_{io\_y}$ denote the iteration time with a total of $x$ GPUs and the I/O time of $y$ GPUs per machine respectively. The speedup of using $N_g$ GPUs can be formulated by
\begin{equation}\label{equ:speedup}
\begin{split}
S &= \cfrac{N_{g}/t_{iter\_N_{g}}}{1/t_{iter\_1}}=N_g\frac{\text{max}\{t_{io\_1}+t_{h2d}, t_{f} + t_{b}\} }{\text{max}\{t_{io\_N_{g}} + t_{h2d}, t_{f} + t_{b} + t_{c}^{no}\}}.
\end{split}
\end{equation}
Therefore, in order to achieve good scalability, one should reduce the overheads of I/O and gradient communications.

Regarding the I/O overhead, all DL frameworks exploit multi-threading to read data and buffer data for GPU computing. However, except Caffe-MPI, the other three frameworks do not use GPU buffers to parallelize the tasks of transferring data from CPU memory to GPU memory. In other words, Caffe-MPI starts the tasks $\mathsf{T}_{40}$-$\mathsf{T}_{43}$ after $\mathsf{T}_{36}$-$\mathsf{T}_{39}$ are finished, while CNTK, MXNet and TensorFlow wait until $\mathsf{T}_{35}$ is finished. Regarding the gradient communication overhead, Caffe-MPI, MXNet and TensorFlow overlap the gradient communication tasks with the back propagation computing tasks, while CNTK dose not do so. For example, task $\mathsf{T}_{32}$ begins after tasks $\mathsf{T}_{20}$-$\mathsf{T}_{23}$ are finished in Caffe-MPI, MXNet and TensorFlow, while it is executed after $\mathsf{T}_{28}$-$\mathsf{T}_{31}$ are finished in CNTK. So, we have $t_{c}^{no}<\sum_{l=1}^{L}t_c^{(l)}$ in Caffe-MPI, MXNet and TensorFlow, but $t_{c}^{no}=\sum_{l=1}^{L}t_c^{(l)}$ in CNTK. The model can also be verified in the section of exeperimental results.

\section{Experiments and Analysis} \label{methods}
In this section, we first introduce the experimental environment, and then we present the experiment design and methods with the purpose of identifying how communication tasks impact the scalability of S-SGD.

\subsection{Experimental environment}
We use two different 4-node GPU clusters for the experiments. Cluster 1 uses the slow intra/inter connections (i.e., PCIe and 10GbE) between Nvidia Tesla K80 GPUs, and Cluster 2 uses the fast intra/inter connections (i.e., NVLink and 100Gbps InfiniBand) between Nvidia Tesla V100 GPUs. The autoboost feature is disabled on all GPUs. Both K80 and V100 GPUs run at their default frequencies (i.e., 562 MHz and 1370 MHz, respectively). Table \ref{table:multigpuetup} shows the hardware setting of the two GPU clusters.
\begin{table}[!ht]
	\begin{threeparttable}
	\centering
	\caption{The experimental hardware setting.}
	\label{table:multigpuetup}
	\begin{tabular}{|l|l|l|}
		\hline
		Hardware& 	Cluster 1 & Cluster 2     \\\hline
		\hline
		GPU (Nvidia) 		& 	Tesla K80 GPU x4  &  Tesla V100 GPU x4  \\\hline
%		Connection & PCIe ($\textasciitilde$15GB/s) & NVLink ($\textasciitilde$95GB/s) \\\hline
		Connection & PCIe (15GB/s) & NVLink (95GB/s) \\\hline
		CPU	(Intel) 		&	Xeon E5-2650v4 Dual & Xeon Gold 6126 CPU Dual\\\hline
		Network		&	10Gbps Ethernet & 100Gbps InfiniBand \\\hline
		Memory		&	256 GB (3.5GB/s) & 256 GB (3.5GB/s)	\\\hline
		Storage system	&	NFS (1.1GB/s) & SSD (367.30MB/s) \\\hline
	\end{tabular}
	\begin{tablenotes}
	\item[]Note: The speed of connections is measured by \textit{p2pBandwidthLatencyTest} from CUDA SDK samples. The memory speed and storage performance are both measured by the $dd$ utility.
	\end{tablenotes}
	\end{threeparttable}
	\vspace{6pt}
\end{table}

The operating system of the K80 GPU cluster is CentOS 7.2 with CUDA-8.0 installed, while the newer GPU V100 cluster is installed with CentOS 7.3 and CUDA-9.0. We choose four state-of-the-art distributed DL frameworks at their current latest release versions (at the time that we did the experiments) for evaluation. Versions of tested frameworks installed in two clusters are shown in Table \ref{table:software_new}. The versions of CUDA related libraries are cuDNN v7.0 and NCCL v2.1.5.

\begin{table}[!ht]
	\centering
	\caption{The softwares used for experiments.}
	\label{table:software_new}
	\begin{tabular}{|l|l|l|}
		\hline
		Software  &  Cluster 1 & Cluster 2  \\
		\hline\hline
		Caffe-MPI      &  2.0	  	& 2.0   \\\hline
		CNTK      &  2.3	  	& 2.4   \\\hline
		MXNet 	  &  1.1.0 	& 1.1.0   \\\hline
		TensorFlow & 1.7		& 1.7         \\\hline
	\end{tabular}
%	\vspace{-10pt}
\end{table}
\subsection{Methodology}
Three popular CNNs (i.e., AlexNet \cite{krizhevsky2012imagenet}, GoogleNet \cite{szegedy2015going} and ResNet-50 \cite{he2015deep}), that are successfully applied on the ILSVRC-2012 ImageNet data set \cite{deng2009imagenet}, are used to do the performance comparison under different software and hardware configurations. The details of the tested CNNs are shown in Table \ref{table:networksetup}. Notice that the machines in Cluster 1 share the data set via NFS, while each machine in Cluster 2 has a complete copy of the data set. The data formats for different frameworks are not the same, and we use the methods proposed in \cite{shi2017performance} to fetch data when running the experiments.
%\begin{itemize}
%	\item Caffe-MPI: The LMDB database is used. Original JPEG images of the data set are converted to LMDB records in advance.
%	\item CNTK: There is no pre-converted data format for CNTK. It needs to read and decompress the original JPEG images during training.
%	\item MXNet: A binary file that contains all the images is used. It also dose not need to decompress the image files.
%	\item TensorFlow: Similar with MXNet, it uses a pre-converted file format called \textit{TFRecord}.
%\end{itemize} 

\begin{table}[htbp]
	\begin{threeparttable}
		\centering
		\caption{The tested deep neural networks.}
		\label{table:networksetup}
		\begin{tabular}{|c|c|c|c|}
			\hline
			Network & Number of Layers & Number of Parameters  & Batch size \\\hline\hline
			AlexNet   & 8     & \textasciitilde 60 millions   & 1024  \\\hline
			GoogleNet & 22    & \textasciitilde 53 millions & 64 \\\hline
			ResNet-50 & 50 	  & \textasciitilde 24 millions & 32 \\\hline
		\end{tabular}
		\begin{tablenotes}
			\item[]Note: The local response normalization (LRN) operation in AlexNet is excluded because it is not supported by CNTK by default. The batch size $B$ specified here is used for a single GPU, and it needs a total of $B\times N_g$ samples per iteration on $N_g$ GPUs.
		\end{tablenotes}
	\end{threeparttable}
%	\vspace{-5pt}
\end{table}

For each experiment, we run more than 100 iterations to calculate the average training time of one mini-batch, and the performance of the training system is represented by the average samples per second.

\subsection{Experimental Results and Analysis} \label{results}
We first present the results of multiple GPUs on a single node to illustrate the impact of the intra-node communications (i.e., PCIe and NVLink), and then present the results of GPU clusters to show the impact of inter-node communications (i.e., 10GbE and 100Gbps InfiniBand). In this paper, we adopt weak scaling, which means the valid mini-batch size scales with the number of GPUs, and each GPU keeps the same number of samples \cite{goyal2017accurate}. The performance is measured by the throughput of the system (i.e., the number of training samples can be processed per second). Ideally, the system performance should be proportional to the number of GPUs.
\begin{figure*}[!h]
	\centering
	\subfigure[The server with K80 GPUs and PCIe]%\label{fig:gradis}
	{
		\includegraphics[width=0.48\linewidth]{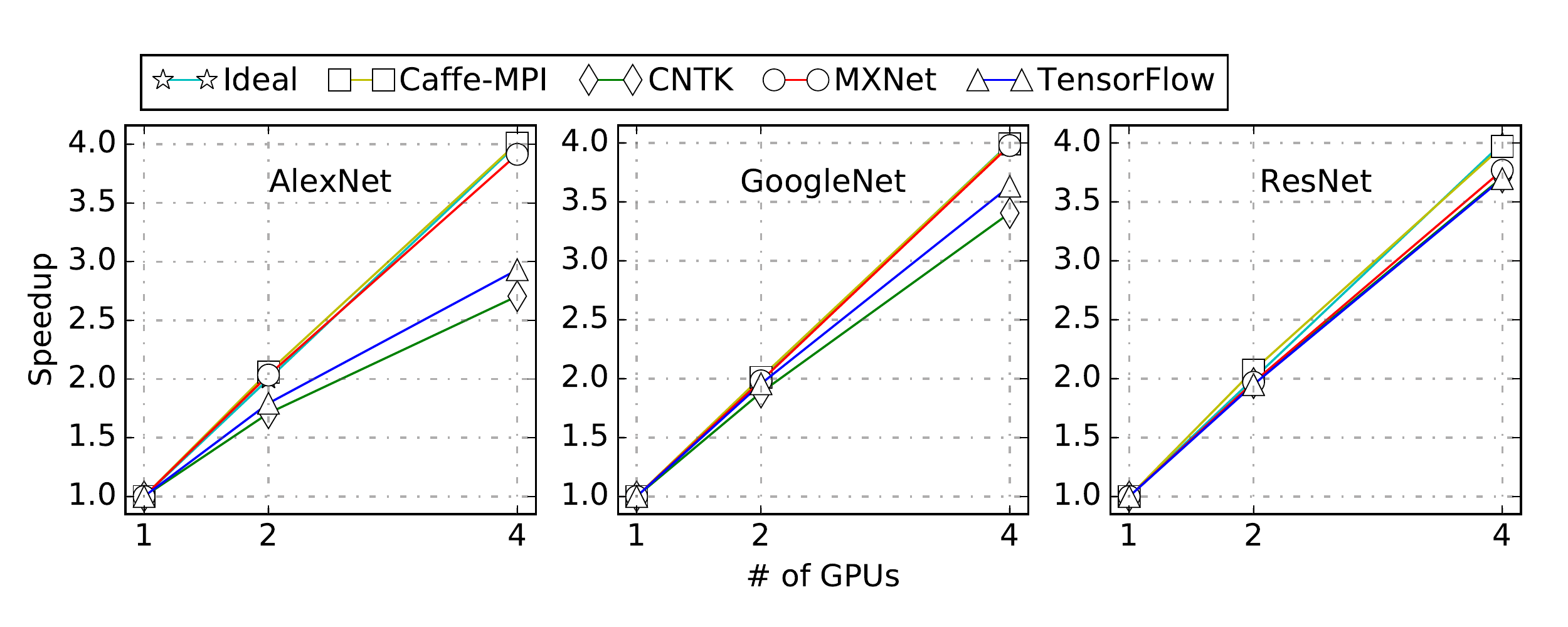}
	}
\vspace{6pt}
	\subfigure[The server with V100 GPUs and NVLink]%\label{fig:commoverhead}
	{
		\includegraphics[width=0.48\linewidth]{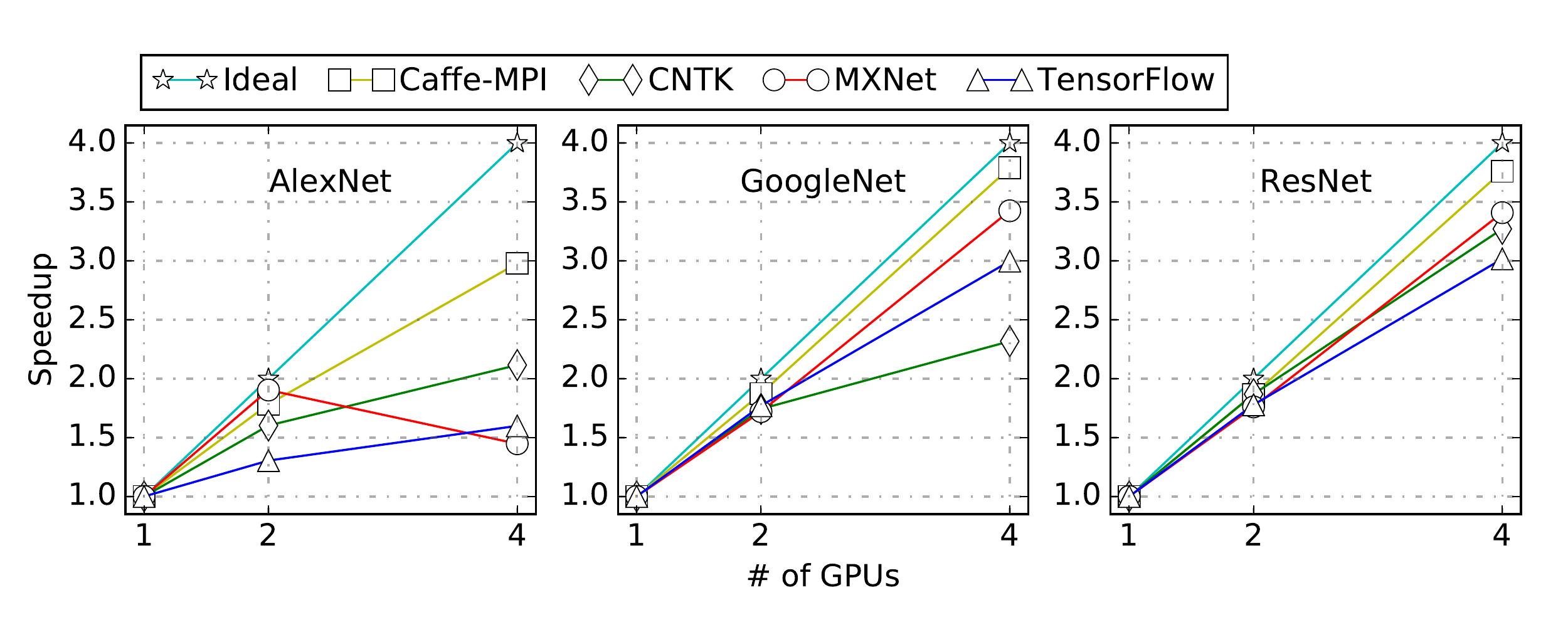}
	}
%		\vspace{-10pt}
	\caption{Scaling performance on a single node.}
	\label{fig:multiplegpus}
%	\vspace{-10pt}
\end{figure*}
\subsubsection{Multiple GPUs on a Single Machine}
Fig. \ref{fig:multiplegpus} shows the scaling performance of four DL frameworks running on a machine with one, two, and four GPUs. The baseline is the performance of a single GPU.

On the K80 server (Fig. \ref{fig:multiplegpus}a), all frameworks achieve good scaling efficiencies (up to 95\%) except that CNTK and TensorFlow don't perform well in AlexNet with 4 GPUs. This is mainly because we use a much larger batch size for AlexNet and the cost of data preprocessing is proportional to the mini-batch size. The data set for Caffe-MPI and MXNet are pre-converted binary formats of input data, and do not need further decoding during the training, while CNTK and TensorFlow need to decode the JPEG files by CPUs before being transferred to GPUs. Since a large number of samples (4096 images per iteration for 4 GPUs) need to be decoded, it takes a relatively long time to decode the data on CPUs compared to the GPU computing tasks, which results in poor scaling efficiency of CNTK and TensorFlow.

On the V100 server (Fig. \ref{fig:multiplegpus}b), the speedup of every framework is worse than that achieved on the K80 server, although the high-speed NVLink is used. From Eq. \ref{equ:speedup}, we can see that the speedup depends on three key factors: the I/O speed, GPU computing performance, and gradient communication performance. Faster computing speed requires faster I/O and communication to maintain a good scaling efficiency. Notice that the K80 GPU has 4.37 TFlops peak computation capability, while the V100 GPU has 125 TFlops peak computation capability with Tensor Cores. Our experiments show that V100 is about 10x faster than K80 in the computing tasks. However, the storage system on the V100 server is about 3x slower than the K80 server, so on the I/O-bound neural network like AlexNet, the scaling efficiency on the V100 server is much worse than that of the K80 server. Regarding GoogleNet and ResNet which require a small number of samples per iteration, the I/O time is negligible; but the gradient communication turns out to limit the scalability because NVLink is only about 6x faster than PCIe.

\begin{figure*}[!ht]
	%	\vspace{-10pt}
	\centering
	\subfigure[The K80 cluster with 10GbE]%\label{fig:gradis}
	{
		\includegraphics[width=0.48\linewidth]{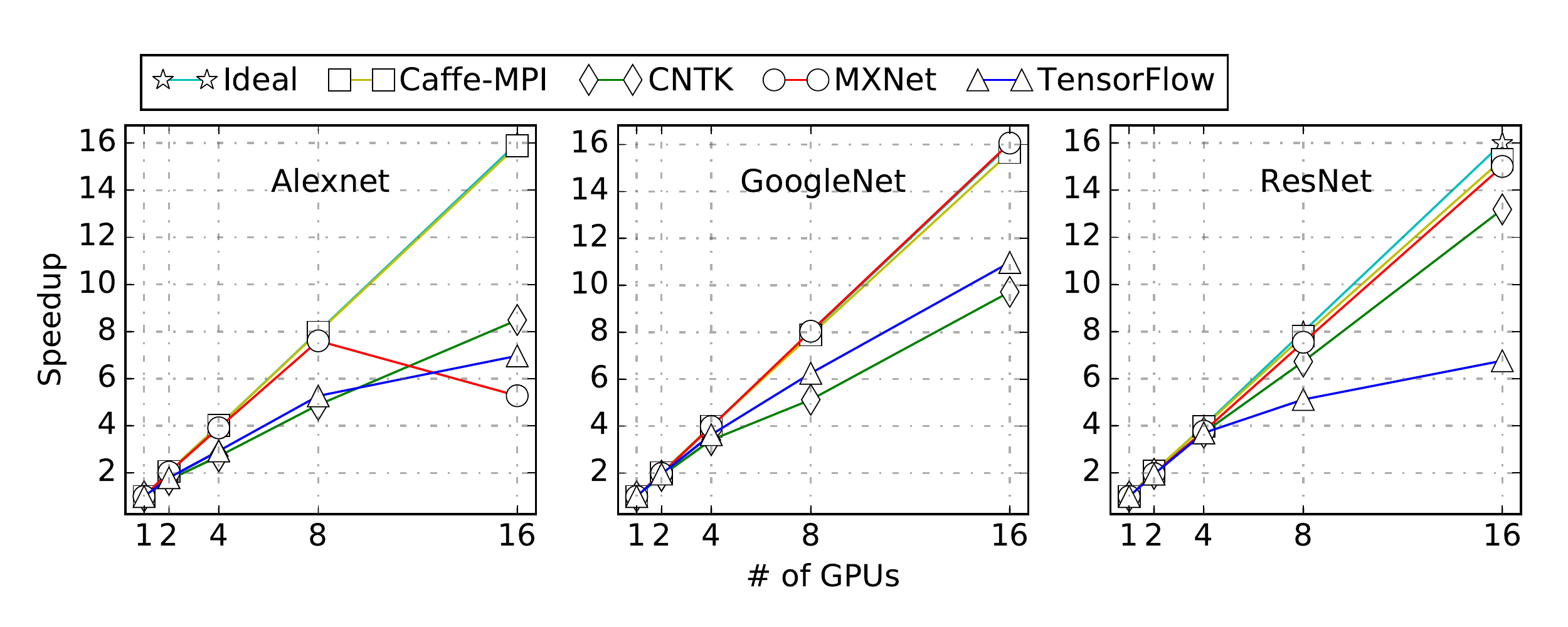}
	}
\vspace{6pt}
	\subfigure[The V100 cluster with 100Gbps InfiniBand]%\label{fig:commoverhead}
	{
		\includegraphics[width=0.48\linewidth]{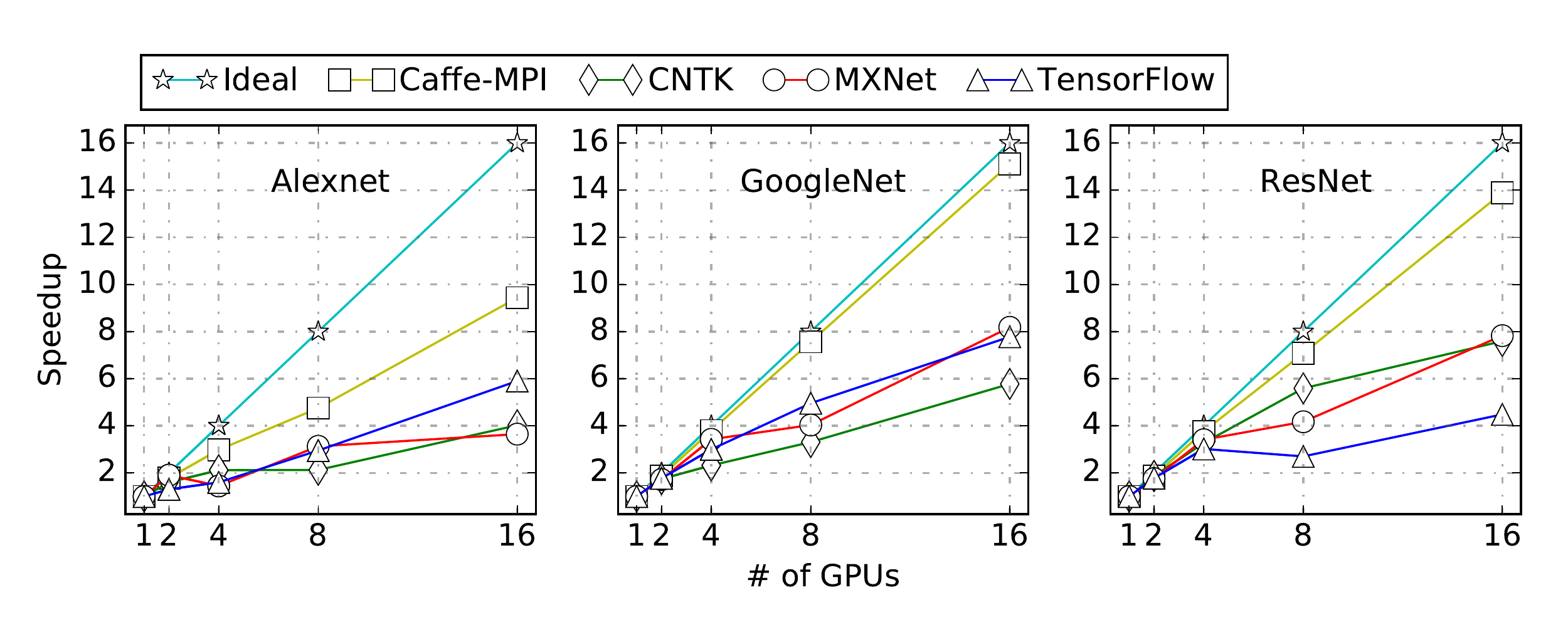}
	}
%			\vspace{-10pt}
	\caption{Scaling performance with multiple machines (each machine has 4 GPUs).}
	\label{fig:multilenodes}
%	\vspace{-10pt}
\end{figure*}
\subsubsection{Multiple GPUs on Multiple Machines}
We show the speedups of multiple machines with up to 16 GPUs in Fig. \ref{fig:multilenodes}. The baseline is the performance of a single server with 4 GPUs.

In general, we can observe that all frameworks scale better on the slow K80 cluster than on the fast V100 cluster. On the K80 cluster, the 10Gbps Ethernet is good enough to serve the gradient communications such that they can be fully overlapped with computing tasks. For example, Caffe-MPI and MXNet achieves nearly linear speedup on GoogleNet and ResNet due to their careful design of gradient aggregations. On AlexNet, however, due to its large mini-batch size and huge amount of model parameters (60 millions), all frameworks don't scale very well on the 4-node cluster. On ResNet, TensorFlow performs the worst mainly because it uses \textit{grpc}\footnote{grpc: https://grpc.io/} for gradient communications which results in relatively high latencies as compared to NCCL2 which is used by Caffe-MPI and CNTK.

On the V100 cluster, it is seen that except Caffe-MPI, the other three frameworks scale poorly across multiple machines. With V100 GPUs, the time cost of computing tasks is significantly reduced, while the overhead of inter-node gradient communication becomes much larger than the case of intra-node (12.5GB/s for InfiniBand vs. 95GB/s for NVLink). As we have already seen in Fig. \ref{fig:multiplegpus} that the fast intra-node communication cannot be totally hidden, so the slower communication through a network results in worse scaling efficiencies. Taking the training of ResNet with Caffe-MPI as an example, the back propagation time on a K80 GPU is about 0.243s, while the overhead of gradient communication is about 0.23s; so the overhead of data communications can be hidden to achieve nearly linear scaling efficiency. However, on the V100 GPUs, the back propagation time is reduced to 0.0625s while the cost of gradient communication is about 0.0797s, so the system becomes communication-bounded. We notice that the communication efficiency on 100Gbps InfiniBand with NCCL2 is only about $9.6\%$ when training ResNet, which indicates a large room of further optimization.

To summarize, we find three main factors to improve the scaling efficiency, i.e., overlapping the pre-fetch of data with computing, overlapping communications with computing, and the efficient data exchanging algorithm. CNTK, MXNet and TensorFlow only implement some of these optimization strategies. Caffe-MPI considers all factors and achieves the best scaling performance, but there is still room for further improvement because even NCCL2 can only achieve $9.6\%$ communication efficiency on the 100Gbps InfiniBand network when training ResNet.
\begin{table*}[!h]
	\centering
	\caption{The measurements for prediction of DAG.}
	\label{table:numbersfordag}
	\begin{tabular}{|l|l|}
		\hline
		Name &  Description \\\cline{1-2}
		\hline
		\hline
		$B_{io}$ & The bandwidth of the hard disk, and it is measured via the $dd$ command \\\cline{1-2}
		$B_{pcie}$ & The bandwidth of PCIe, and it is measured via the CUDA SDK\\\cline{1-2}
		$D$ & The data size of the input data, which is related to the mini-batch size\\\cline{1-2}
		$t_{io}$ & The time of I/O is calculated by the division of the input data size and $B_{io}$ \\\cline{1-2}
		$t_{h2d}$ & The time of data transfer from the CPU side to the GPU side is measured by the division of the input data size and $B_{pcie}$ \\\cline{1-2}
		$t_f^{(l)}$ & The layer-wise feed-forward time is measured from Caffe-MPI which invokes the cuDNN library \\\cline{1-2}
		$t_b^{(l)}$ & The layer-wise backward propagation time is measured from Caffe-MPI which invokes the cuDNN library \\\cline{1-2}
		$t_c^{(l)}$ & The layer-wise gradient communication time is measured from Caffe-MPI which invokes the NCCL2 library\\\cline{1-2}
	\end{tabular}
	\vspace{10pt}
\end{table*}

\subsection{Accuracy of the DAG model}

\begin{figure}[!h]
	%	\vspace{-8pt}
	\centering
	\includegraphics[width=\linewidth]{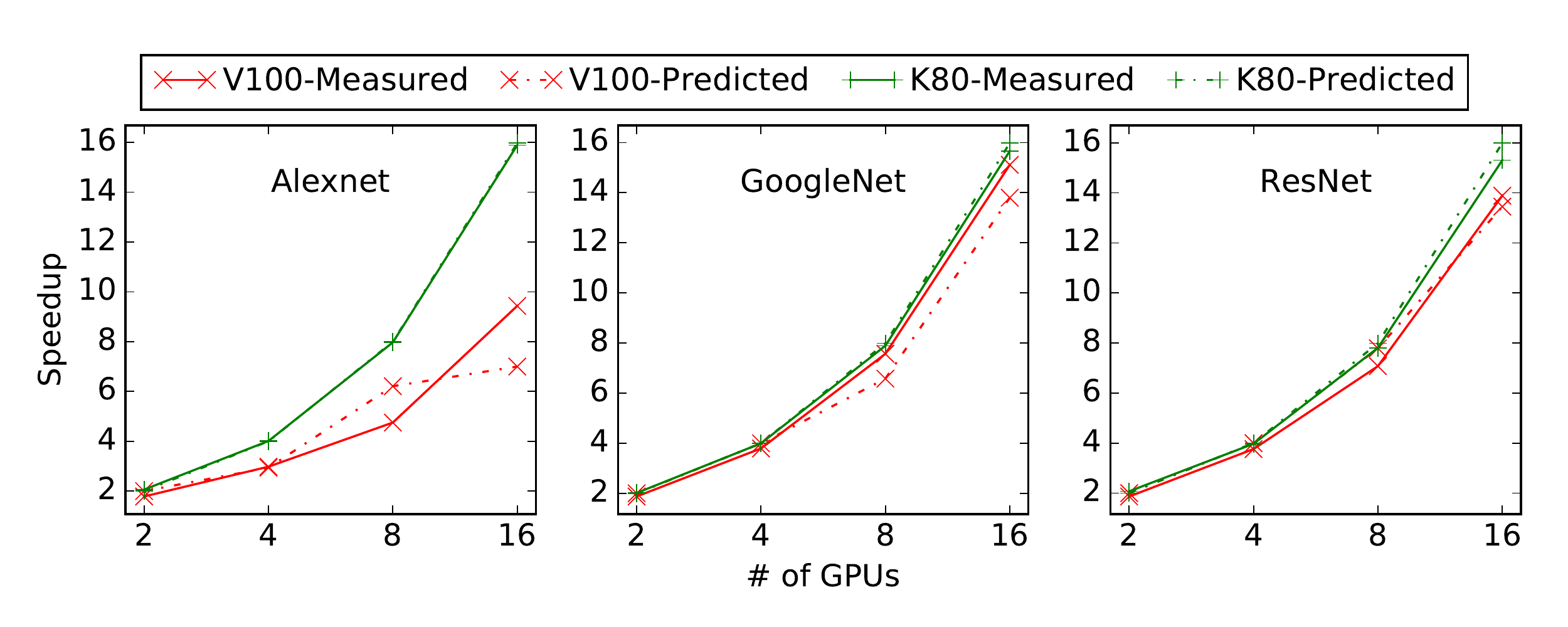}
	%	\vspace{-15pt}
	\caption{Comparison of DAG-based prediction and measurement results on the K80 cluster and the V100 cluster. }
	\label{fig:predict}
	\vspace{10pt}
\end{figure}

In this section, we demonstrate the accuracy of the DAG model by comparing the predicted speedup with the experimental results of Caffe-MPI. 

To do the prediction of the time performance of DAG, we need to measure the numbers from the evaluated CNNs on the specific platforms. The measurement details from the two clusters for prediction are shown in Table \ref{table:numbersfordag}.

Using the known time of each operation in Fig. \ref{fig:dag} on two main platforms (i.e., the V100 GPU cluster connected with 100Gbps InfiniBand and the K80 GPU cluster conncected with 10Gbps Ethernet), we can predict the average iteration time when training AlexNet, GoogleNet and ResNet. The comparison between the prediction with DAG and measurements by Caffe-MPI is shown in Fig. \ref{fig:predict}. The average prediction errors are 9.4\%, 4.7\%, and 4.6\% on AlexNet, GoogleNet and ResNet respectively. Using AlexNet, it is obvious that the speedup over multiple GPUs is hard to be linear on the fast V100 GPUs. The reason is that even using the optimal DAG scheduling, the communication time of gradients cannot be hidden by the computation time. 

The DAG model can serve as a fundamental tool for performance optimization and task scheduling.

\section{Layer-wise trace data set}\label{dataset}
\begin{table}[!ht]
	\centering
	\caption{An example of the trace data with one iteration of AlexNet on the K80 GPU.}
	\label{table:tracesample}
	\begin{tabular}{|l|l|l|l|l|l|}
		\hline
		Id& Name & Forward & Backward & Comm.& Size \\\cline{1-6}
		\hline
		\hline
		0& data& 1.20e+06& 0& 0& 0\\\cline{1-6}
		1& conv1& 3.27e+06& 288202& 123.424& 139776\\\cline{1-6}
		2& relu1& 17234.5& 27650.9& 0& 0\\\cline{1-6}
		3& pool1& 32175.7& 60732.6& 0& 0\\\cline{1-6}
		4& conv2& 3.14e+06& 1.03216e+06& 292.032& 1229824\\\cline{1-6}
		5& relu2& 11507.5& 18422.5& 0& 0\\\cline{1-6}
		6& pool2& 19831.2& 32459& 0& 0\\\cline{1-6}
		7& conv3& 3.886e+06& 791825& 288214& 3540480\\\cline{1-6}
		8& relu3& 4770.3& 10996.3& 0& 0\\\cline{1-6}
		9& conv4& 1.87e+06& 510405& 1.03218e+06& 2655744\\\cline{1-6}
		10& relu4& 4760.26& 7872.45& 0& 0\\\cline{1-6}
		11& conv5& 1.13e+06& 306129& 275772& 1770496\\\cline{1-6}
		12& relu5& 3201.22& 4939.42& 0& 0\\\cline{1-6}
		13& pool5& 5812& 18666.2& 0& 0\\\cline{1-6}
		14& fc6& 44689.7& 73935& 311170& 151011328\\\cline{1-6}
		15& relu6& 295.168& 1092.83& 0& 0\\\cline{1-6}
		16& drop6& 359.744& 131247& 0& 0\\\cline{1-6}
		17& fc7& 19787.8& 34423.8& 610376& 67125248\\\cline{1-6}
		18& relu7& 295.04& 451.904& 0& 0\\\cline{1-6}
		19& drop7& 358.048& 317.312& 0& 0\\\cline{1-6}
		20& fc8& 8033.12& 9922.72& 130964& 16388000\\\cline{1-6}
		21& loss& 1723.49& 293.024& 0& 0\\\cline{1-6}
		
	\end{tabular}
	%	\vspace{-10pt}
\end{table}

We make the trace data set from Caffe-MPI publicly available\footnote{Download address: \url{http://dlbench.comp.hkbu.edu.hk/s/data/traces.zip}}, which could be used for further simulation studies (e.g., tasks scheduling and communication optimizations) for those who do not have access to the expensive GPUs. The trace data set includes the layer-wise time cost of the evaluated three types of CNNs (i.e., AlexNet, GoogleNet and ResNet-50) on the V100 GPU cluster and the K80 GPU cluster. Each record contains the time of feed forward, back propagation, intra/inter-node communication and the size of gradients in an iteration. 

Each trace file contains 100 iterations of the layer-wise time speed. One can use the average time for more accurate measurements. An example with one iteration of AlexNet on two K80 GPUs is shown in Table \ref{table:tracesample}. There are 22 layers including the data layer and some non-learnable layers like activations. Each line indicates the time performance of one layer of that CNN. The meaning of each column in the trace file is as follows:
\begin{itemize}
	\item The first column indicates the layer id;
	\item The second column indicates the pre-defined name of that layer;
	\item The third column is the feed-forward time in microsecond of that layer;
	\item The fourth column is the backward propagation time in microsecond;
	\item The fifth column is the gradient communication time in microsecond, and zero values of some layers indicate that layers are not learnable (i.e., no need to exchange gradients);
	\item The sixth column is the size of gradients that need to be exchanged among GPUs in bytes, and it is the same as the size of model parameters of that layer.
\end{itemize} 
%The first column is the layer id; the second column is the pre-defined name of that layer; the third column is the feed-forward time of that layer; the fourth column is the backward propagation time; the fifth column is the gradient communication time, and zero values of some layers indicate that layers are not learnable (no need to exchange gradients);  The time recorded is all in microsecond.

%\begin{figure}[!h]
%	%	\vspace{-8pt}
%	\centering
%	\includegraphics[width=\linewidth]{tracesample.png}
%	%	\vspace{-15pt}
%	\caption{The trace data of one iteration of AlexNet on the K80 GPU.}
%	\label{fig:tracesample}
%	%	\vspace{-10pt}
%\end{figure}
\vspace{10pt}
\section{Conclusion and Future work} \label{conclusionandfuturework}

This work aims to understand the impact of data communication techniques on the distributed training performance of deep neural networks. We first propose a DAG model to describe the workflow of synchronized SGD in deep learning. Through the DAG model, we identify that the communication tasks (including the I/O tasks) could affect the scaling efficiency of the system. We then conduct extensive empirical studies on the performance of four state-of-the-art DL frameworks (Caffe-MPI, CNTK, MXNet and TensorFlow) by training three DNNs (AlexNet, GoogleNet and ResNet-50) across multiple GPUs and multiple machines. According to our experimental results and analysis, we show some performance gaps among four distribute DL frameworks due to their different optimization strategies. We also show that even the most advanced NVLink and InfiniBand techniques cannot catch up with the fast growth of GPU computing power. This demands for more research efforts from the data communication and networking community to address the communication issues in deep learning.

We will further optimize the pipeline between gradient exchange operations and backward propagation operations to achieve better effective bandwidth since current implementations have no good utilization of network resources.

%\section{Acknowledgements} \label{ack}
%We would like to thank Inspur (Beijing) Co., Ltd for providing the GPU cluster for experiments.

\bibliographystyle{IEEEtran}
\Urlmuskip=0mu plus 1mu
\bibliography{dist-dlbench-icpads2018}
\end{document}